# Negative differential resistance and magnetoresistance in zigzag borophene nanoribbons

Jiayi Liu [a], Changpeng Chen [a, b,*], Lu Han[a], Ziqing Zhu[a], Jinping Wu[b,*]
a. School of science, Wuhan University of Technology, Wuhan 430070 PRChina
b. Material Science and Chemistry Engineering College, China University of Geosciences, Wuhan, 430074, PRChina

Abstract: We investigate the transport properties of pristine zigzag-edged borophene nanoribbons (ZBNRs) of different widths, using the fist-principles calculations. We choose ZBNRs with widths of 5 and 6 as odd and even widths. The differences of the quantum transport properties are found, where even-N BNRs and odd-N BNRs have different current-voltage relationships. Moreover, the negative differential resistance (NDR) can be observed within certain bias range in 5-ZBNR, while 6-ZBNR behaves as metal whose current rises with the increase of the voltage. The spin filter effect of 36% can be revealed when the two electrodes have opposite magnetization direction. Furthermore, the magnetoresistance effect appears to be in even-N ZBNRs, and the maximum value can reach 70%.

1. Introduction

   Over the last decades the field of two-dimensional (2D) atomic-layer systems, including graphene [1–3], transition metal dichalcogenides [4,5], silicone [6,7] and germanane [8] has seen important developments ,both in fundamental aspects and prospective technological applications. [10–17]
   Recently, a new 2D material, namely, borophene, has been grown successfully on single crystal Ag(111) substrates under ultrahigh-vacuum conditions and immediately received considerable attention due to their extraordinary properties[18]. Theoretical studies have proposed various structures for borophene[19–28], while investigations by scanning tunneling microscopy have shown that borophene has planar structure with anisotropic corrugation. It is metallic with highly anisotropic electronic properties. On this foundation, Bo Peng et al [28] investigated electronic, optical and thermodynamic properties of borophene by first-principles calculations. Jianhui Yuanet et al [29] found Young's moduli of the ABNRs (armchair borophene nanoribbons) are slightly greater than those of the ZBNRs(zigzag borophene nanoribbons) using the molecular dynamics simulations.
   As is well known, quasi-one dimensional C, Si, P nanoribbon have been studied theoretically and experimentally due to its unique properties [30-35]. Moreover, symmetry-dependent and whether a mirror plane σ exists strongly determine the transport behaviors of 2D materials, i.e., graphene nanoribbons (ZGNRs), silicone nanoribbons (ZSiNRs) and germanene nanoribbons (ZGeNRs) [36-38]. Do pristine zigzag borophene nanoribbons (ZBNRs) also exhibit symmetry-dependent transport properties? To explore this question and find out whether ZBNRs have magnetoresistance or the spin-filtering effect in ferromagnetic, we investigate the transport properties of borophene by means of the density functional theory (DFT) and the nonequilibrium Green's function (NEGF) method in this paper. We find that they still show symmetry-dependent transport properties since the effect is not strong enough like above investigated materials. Magnetoresistance and spin filtering effect can also be observed in ZBNRs.

2. Models and Calculation Method

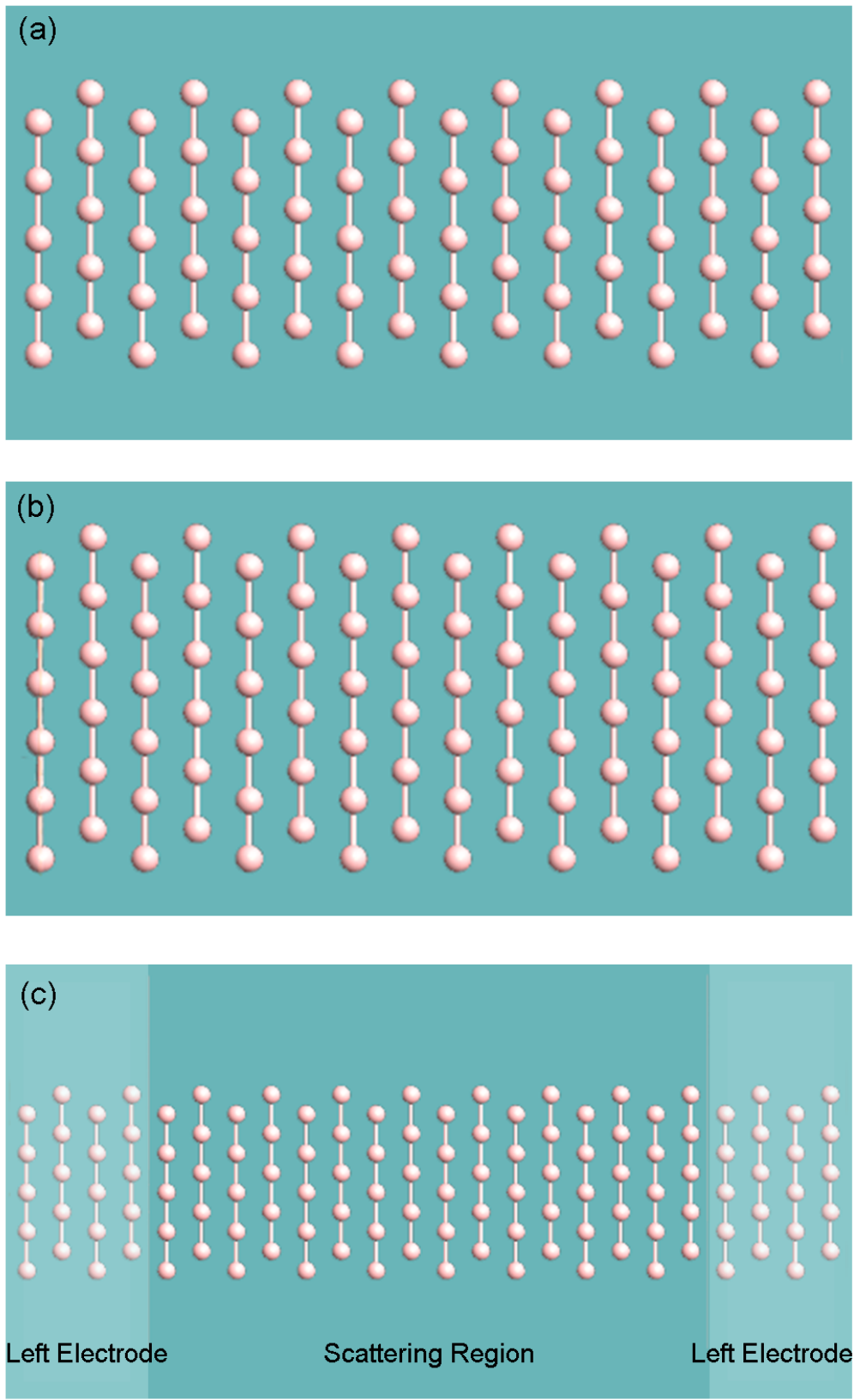

Fig.1. (a) and (b) Structure of 5-ZBNR (a) and 6-ZBNR. (c) Schematic of the two probe system.

To compare the difference in transport properties of pristine odd-N ZBNRs and even-N ZBNRs, we picked up with the widths of 5 and 6,which are noted here as 5-ZBNR and 6-ZBNR,respectively. As shown in Fig.1, the optimized geometric structure of zigzag BNRs is classified by the width of the borophene nanoribbons. The models are optimized and all electronic transport properties are carried out using density-functional theory (DFT) calculations with the local spin density approximation (LSDA) for exchange correlation functional within the generalized gradient approximation (GGA) and as implemented in the Atomisix Tool Kit-VirtualNanoLab (ATK-VNL), which is based on the nonequilibrium Green functions. Structural optimizations are carried out until the absolute value of atomic forces less than 0.01eV/A .For transport calculation, k-point grid 1x1x110 were used. Furthermore, we set the grid mesh cutoff of 150Ry for all calculations to ensure better accuracy. The quantum current is calculated according to the Landauer–Büttiker formula:

$$I(V) = \frac{e}{\hbar} \int_{\mu_L(V_b)}^{\mu_R(V_b)} T(E,V)[f_L(E,V) - f_R(E,V)]dE$$

where is the Fermi-Dirac distribution for the left (L) and right (R) electrodes, (σ=L/R) is the electrochemical potential of the left/right electrode (,and T(E,V) is the transmission coefficient that depends on the energy E and bias voltage V, defined as

$$T(E,V) = Tr[\Gamma_L(E,V)G^R(E,V)\Gamma_R(E,V)G^A(E,V)]$$

where is the retarded (R) and advanced (A) Green function, is the imaginary parts of the right (R) and left (L) self-energies.

3. Results and Discussion

We calculated the I-V characteristics of 5-ZBNR and 6-ZBNR to study whether ZBNRs would exhibit symmetry-dependent transport properties as ZGeNRs and ZSiNRs, and the results are given in Fig.2(a). It is obvious that 5-ZBNR and 6-ZBNR show different I-V relationships as the current through 5-ZBNR is much higher than that of through 6-ZBNR under the same applied bias. This means ZBNRs show similar symmetry-dependent transport properties to hydrogenated ZSiNRs and ZGeNRs[37,38].

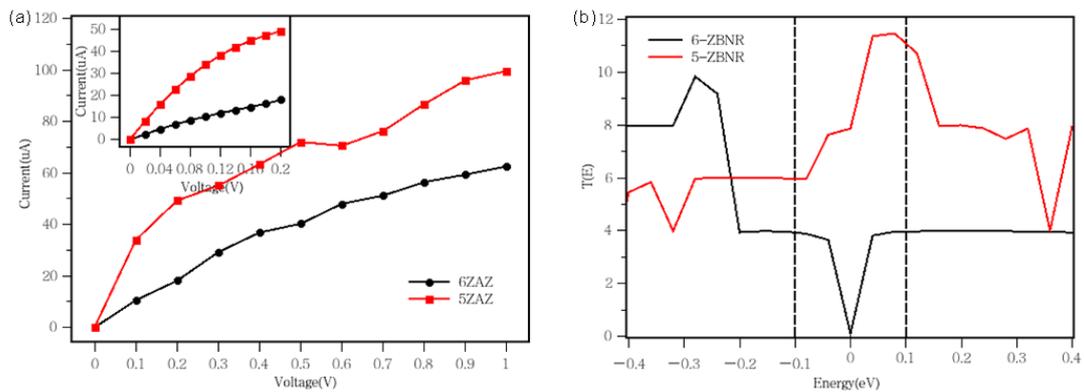

Fig.2. (a) The current as functions of the applied bias for 5-ZBNR and 6-ZBNR. The inset shows the local expansion for lower bias range. (b) The transmission spectra of 5-ZBNR and 6-ZBNR under 0.1

bias voltage. The region between the two vertical black dash lines is the bias window.

The transmission spectrum T(E,V) of 5-ZBNR and 6-ZBNR are calculated under 0.1V in Fig.2(b). The two vertical black dash lines show bias window. One can see easily that there are strong transmissions for 5-ZBNR. However, from the transmission spectrum of 6-ZBNR, it can be seen that the transmission coefficient is nearly zero around the Fermi level. Compared with 5-ZBNR, 6-ZBNR presents remarkably weaker transmissions under 0.1V, which is in accordance with the calculated I-V curve.

In addition, the current of 5-ZBNR shows slight negative differential resistance(NDR) behavior in the bias range between 0.5 V and 0.6 V. The NDR effect had aroused much interest since it is crucial for several electronic components such as resonant tunneling diode. Besides, the geometrical factors such as the ribbon width variation and symmetry play an important role in NDR.

To understand the mechanism of I-V characteristics difference and NDR, we calculate the transmission spectrum for 5-ZBNR at finite biases from 0 to 1V and plot the result in Fig.3. We find that there are higher transmission coefficients in bias window around zero bias, which contributes to the current increasing at the lower bias range. With the voltage increases, the transmission comes into bias windows which lead to the increasing of current. But the color in bias window becomes cool from red, which shows the value of the transmission coefficient is smaller. Therefore, the current increases slowly from 0.2 to 0.5V. However, the total value of transmission coefficient coming into the bias window drops further at the bias range from 0.5V to 0.6V. So, NDR effect is observed for 5-ZBNR system.

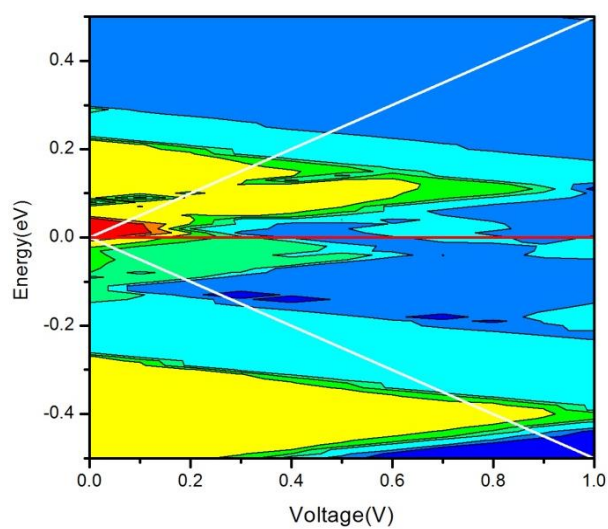

Fig.3. The total transmission spectrum as a function of the bias voltage and electron energy for 5-ZBNR. The red indicates the largest value and black means almost zero. The horizontal red line shows the average Fermi level and the two gradient white line show bias window.

In order to further explain the origin of the NDR phenomenon appearing in 5-ZBNR system, we calculate the bias voltage-dependent MPSH eigenvalue and eigenstate of scattering region combined with the transmission spectrum under 0.6V. Fig.4 exhibits the

MPSH of the highest occupied molecular orbital (HOMO), the lowest unoccupied molecular orbital (LUMO), and their nearby orbitals HOMO-1 and LUMO+1. From this figure, we can obviously find that HOMO, LUMO and LUMO+1 are nearly localized for 5-ZBNR at 0.6V, while those at 0.5V is delocalized. The delocalized orbits across the molecule will greatly enhance the probability of an electron going through the molecule.

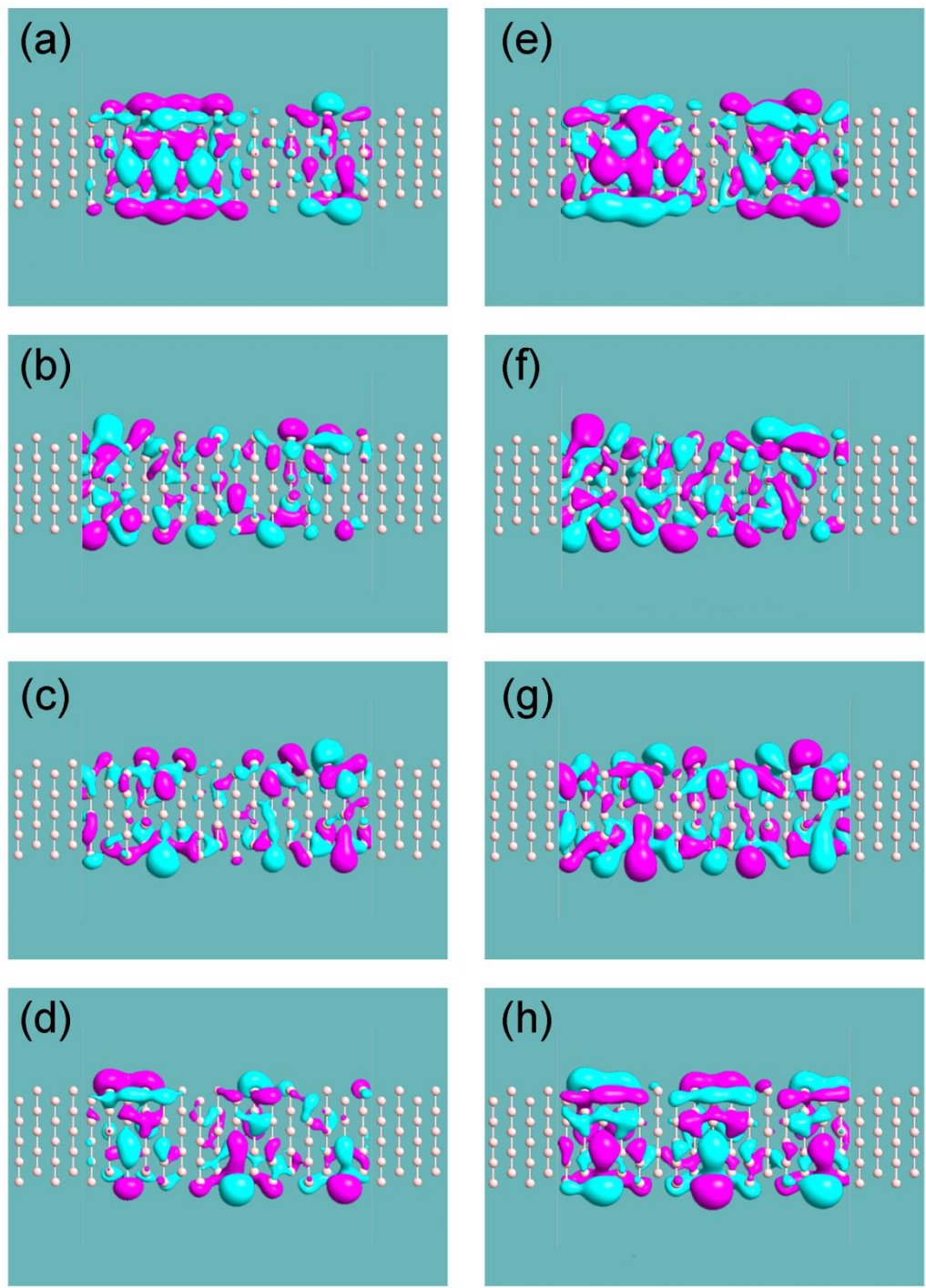

Fig.4. (a)-(d) Spatial distribution of MPSH LUMO+1,LUMO,HOMO and HOMO-1 respectively for 5-ZBNR under 0.6V. (e)-(f) The same as (a)-(d) but under 0.5V.

The above results show that ZBNR also exhibit symmetry-dependent transport properties which is the same as ZGeNR and ZsiNR. Previous study has shown that magnetoresistance(MR) and the spin-filtering effect(SFE) can be found in even-N system of ZGeNRs and ZSiNRs, considering the dependence of the transport properties on the width effect [37,38]. To further explore whether even-N ZBNR has magnetoresistance or the spin-filtering effect, we construct a two probe system of the left and right electrodes with two spin configurations: (i) P (parallel) configurations, which the two electrodes have the same magnetization direction; and (ii) AP (antiparallel) configuration, which the two electrodes have opposite magnetization direction. We obtain these two configurations by applying external magnetic field. The spin density of the two configuration of 6-ZBNR under zero bias is presented in Fig.5. It can be found that the main magnetic distributions are around the edge B atoms. Applying a bias between the two electrodes, we calculate the corresponding current in Fig.6. The current of the different spin components under different configuration is similar in the bias range of [0V, 1V], which does not exhibit the great magnetoresistance. But in lower bias range of [0V, 0.2V], one can easily see that the currents of different circumstance are separated.

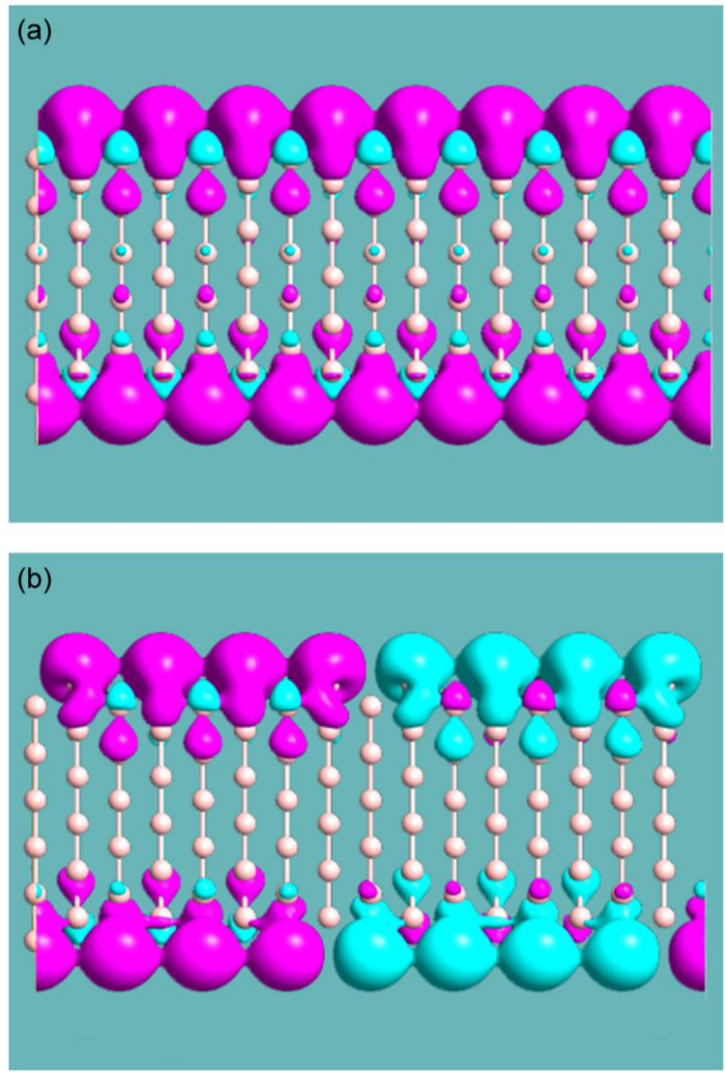

Fig.5. (a) and (b) Spin density of 6-ZBNR with P and AP spin configurations, respectively.

For assessing the spin polarization, the bias-dependent spin filter efficiency (BDSFE) for the AP configuration at finite bias is plotted by using the formula BDSFE = ($I_{up}$ - $I_{down}$) / ($I_{up}$ + $I_{down}$) in Fig.7. It can be clearly seen that the maximum values of BDSFE reaches above 36% at the bias of 0.02V, whereas the spin filter efficiency dependent on the applied bias continues to decline with the increase of the voltage. It indicates even 6-ZBNR shows a weak spin-filter effect in AP magnetism configuration, compared with the case of 6-ZGeNR and 6-ZSiNR [37,38].

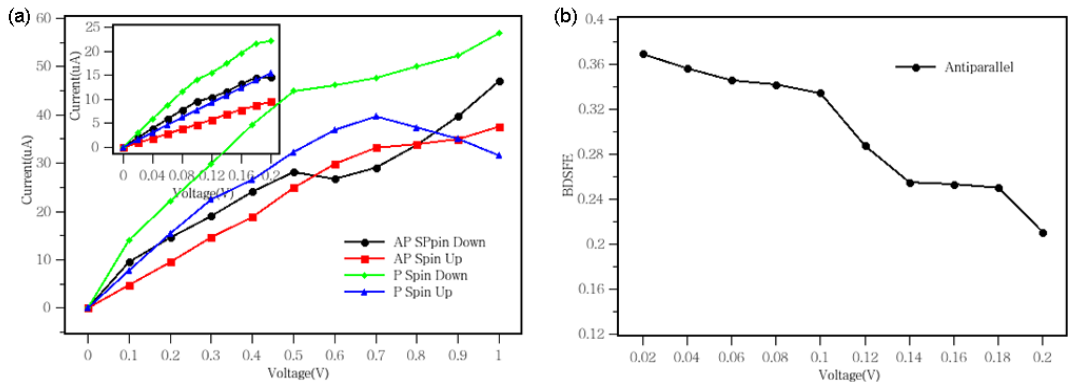

Fig.6. (a) The total current I of 6-ZBNR with different electronic configurations. The inset shows the local expansion for lower bias range.

The interesting spin transport properties can be understood from the transmission spectrum shown in Fig.7. There are two significant transmission peaks, i.e., $P_1$, $P_2$, in bias window under the bias of 0.1V. Note that only the peaks in the bias window would affect the electron transport [39]. The graph shows that $P_1$ produces a relatively large $I_{down}$, while the amplitude of $P_2$ is lower than $P_1$ resulting in that $I_{down}$ is larger than $I_{up}$. Therefore, a SFE appears. When the bias increases to 0.14V, $P_1$ becomes smoother around the Fermi level but the amplitude is nearly unchanged. In contrast, the amplitude of $P_2$ becomes higher. Because of that, $I_{up}$ rise more quickly than $I_{down}$, leading to the BDSFE decreasing from 0.1V to 0.14V.

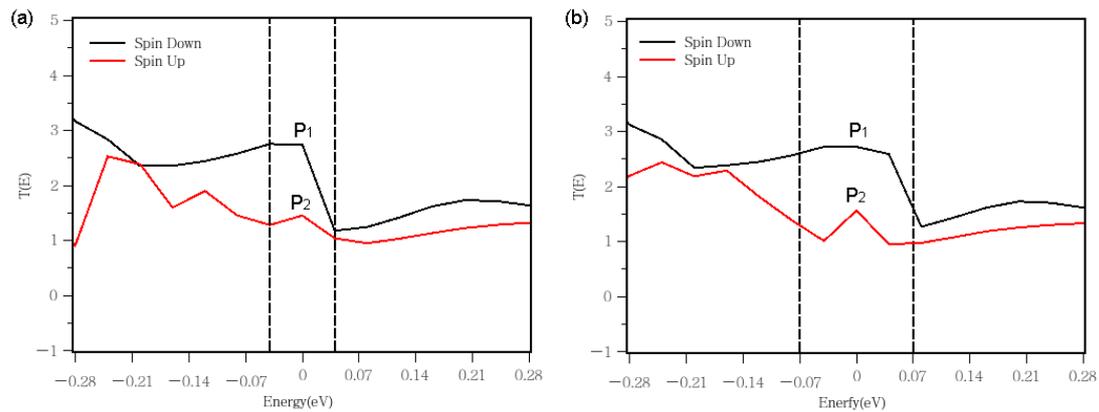

Fig.7. (a) and (b) The spin-polarized transmission spectra of 6-ZBNR under 0.8V and 1.4V, respectively. The region between the two vertical black dash lines is the bias window.

In Fig.8, we plot the I-V curve at the bias range of [0V, 0.2V] and draw the magnitude of bias-dependent MR according the formula MR=($I_P$-$I_{AP}$)/$I_{AP}$ to study the magnetoresistance more intuitively ($I_P$ and $I_{AP}$ are current in P and AP configurations, respectively. Compared with 6-ZSiNR holding MR in the order of 100000% [31], 6-ZBNR has the weaker MR, whose spin-up, spin-down and total MR are all higher than 45%. Especially for spin-up, MR is over 55%, even the maximum value of MR reaches 70%. Moreover, MR is stable under the variation of bias.

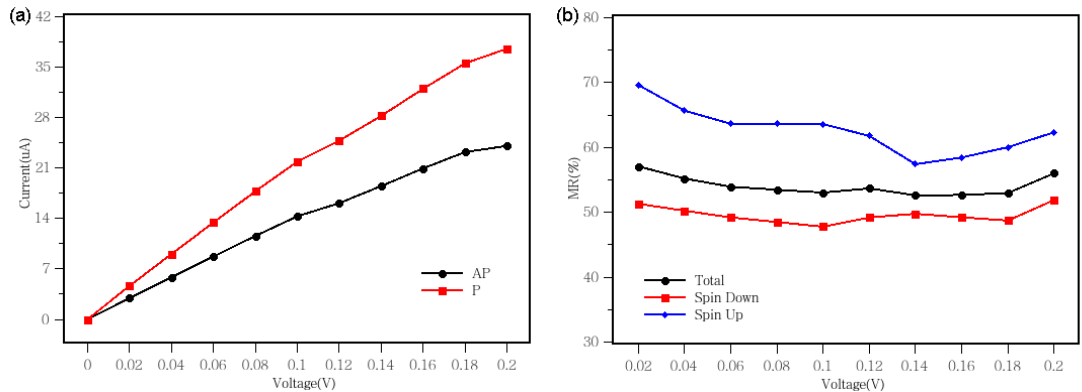

Fig.8 (a) The I-V curves for the P and AP configurations.  (b) The spin up, spin down and total magnetoresistance.

To explain the mechanism of the MR effect of 6-ZBNR, we depict the main transmission eigenchannels at Fermi level under zero bias in Fig.9. We can apparently notice that spin-up and spin-down electrons have different transmission channels. In the P configuration, the spin-up eigenchannel is delocalized, while the spin-down one is quasi-localized, which contributes to the larger current of P configuration. In contrast, the spin-up and spin-down eigenchannel are both localized when the magnetization directions of two electrodes are opposite, which weakens the conductance of the system. Finally, the difference of opening channels leads to the MR effect.

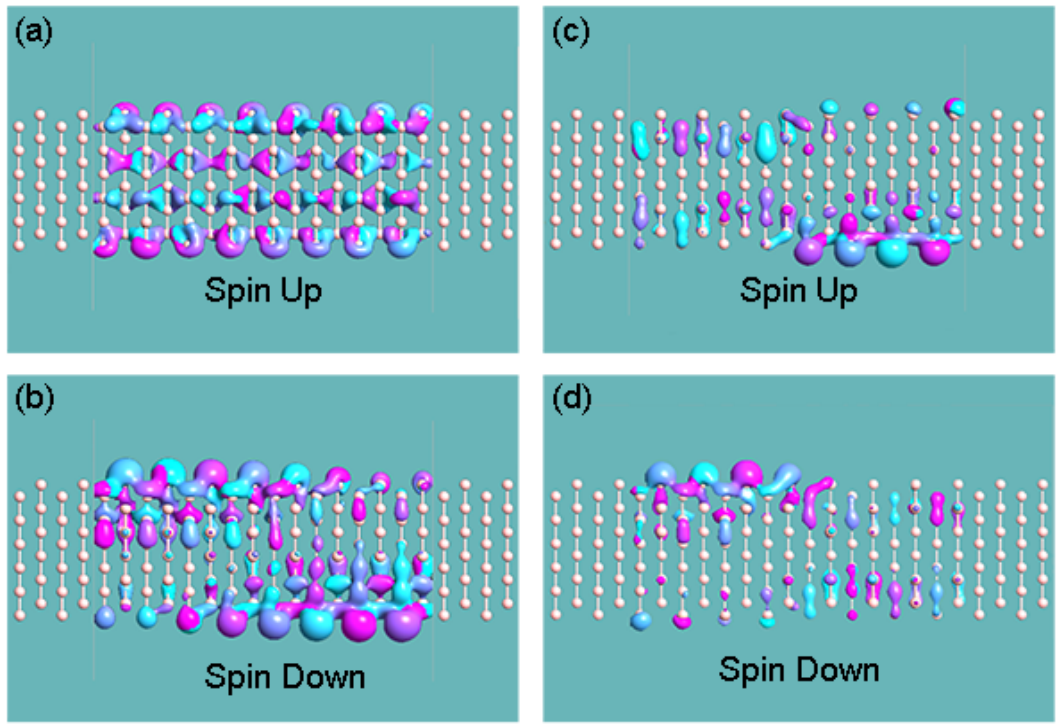

Fig.9 (a) and (b)The main transmission eigenchannel of 6-ZBNR at Fermi level under zero bias when the magenetization directions of two electrodes are parallel. (c) and (d) The same as (a) and (d) but when the two electrodes have different magnetic configuration.

4. Conclusion

In conclusion, ZBNRs of different widths show completely different transport behaviors. Symmetric ZBNRs show the metal transport linear current-voltage dependence, but its current is small, since asymmetric ZBNRs have NDR which can be used in design of electronic applications based on ZBNRs. Moreover, it is found that even-N ZBNRs exhibit magnetoresistance and the spin-filtering effect. When the system is in AP magnetism configuration, the spin-filtering effect can be observed with nearly 36% at certain bias voltage, and the current is relatively small, which result in magnetoresistance. The order of the corresponding magnetoresistance can reach 70%.

**Acknowledgements**
   The authors would like to acknowledge the support by the Project 61177076 supported by National Natural Science Foundation of China and 2016 china national college students' innovative and entrepreneurial training program funding projects.

1. Novoselov, K. S. et al.  Nature 438, 197–200 (2005).
2. Geim, A. K. & Novoselov, K. S. Nat. Mater. 6, 183–191(2007).
3. Geim, A. K. & Grigorieva, I. V. Nature 499,3419–425 (2013).
4. Wang, Q. H. et al. Nat. Nanotech. 7, 699–712 (2012).
5. Fivaz, R. & Mooser, E. Phys. Rev. 163, 743–755 (1967).
6. Vogt, P. et al. Phy. Rev. Lett. 108, 155501 (2012).

7. Houssa, M. et al. Appl. Phys. Lett. 98, 223107 (2011).
8. Bianco, E. et al. ACS Nano 7, 4414–4421 (2013).
9. Majidi, D. and R. Faez. Physica E: Low-dimensional Systems and Nanostructures 86: 175-183 (2017).
10. Berger, C. et al. J. Phys. Chem. B 108,19912–19916 (2004).
11. Liao, L. et al. Nature 467, 305–308 (2010).
12. Schwierz, F. Nat. Nanotechnol. 5, 487–496 (2010).
13. Radisavljevic, B. et al. Nat. Nanotech. 6,147–150 (2011).
14. Wang, H. et al. Nano Lett.12, 4674–4680 (2012).
15. Peng L, Yao K, Wu R, et al. Physical Chemistry Chemical Physics, 17(15): 10074-10079 (2015).
16. Zha D, Chen C, Wu J. Solid State Communications, 219: 21-24 (2015).
17. Zha D, Chen C, Wu J, et al. International Journal of Modern Physics B, 29(09): 1550061 (2015).
18. A.J.Mannix, X.-F. Zhou, B. Kiraly, J. D. Wood, D. Alducin, B. D. Myers, X. Liu, B. L. Fisher,U. Santiago, J. R. Guest, M. J. Yacaman, A. Ponce, A. R. Oganov, M. C. Hersam, and N. P.Guisinger, Science 350(6267): 1513-1516 (2015).
19. I. Boustani, Phys. Rev. B 55, 16426 (1997).
20. H. Tang and S. Ismail-Beigi, Phys. Rev. Lett. 99, 115501 (2007).
21. K. C. Lau and R. Pandey, The Journal of Physical Chemistry C, J. Phys. Chem. C 111, 2906(2007).
22. H. Liu, J. Gao, and J. Zhao, Scientific Reports 3, 3238 (2013).
23. Y. Liu, E. S. Penev, and B. I. Yakobson, Angew. Chem. Int. Ed. 52, 3156 (2013).
24. X.-F. Zhou, X. Dong, A. R. Oganov, Q. Zhu, Y. Tian, and H.-T. Wang, Phys. Rev. Lett. 112,085502 (2014).
25. X.-B. Li, S.-Y. Xie, H. Zheng, W. Q. Tian, and H.-B. Sun, Nanoscale 7, 18863 (2015).
26. J. Yuan, L. W. Zhang, and K. M. Liew, RSC Adv. 5, 74399 (2015).
27. X. Zhang, D. Sun, Y. Li, G.-H. Lee, X. Cui, D. Chenet, Y. You, T. F. Heinz, and J. C. Hone,ACS Applied Materials & Interfaces, ACS Appl. Mater. Interfaces 7, 25923 (2015).
28. Peng B, Zhang H, Shao H, et al. J. Mater. Chem. C 4(16): 3592-3598 (2016).
29. Yuan J, Zhang L W, Liew K M. RSC Advances, 5(91): 74399-74407 (2015).
30. Son Y W, Cohen M L, Louie S G. Nature, 444(7117): 347-349 (2006).
31. Yan Q, Huang B, Yu J, et al. Nano letters, 7(6): 1469-1473 (2007).
32. Xu C, Luo G, Liu Q, et al. Nanoscale, 4(10): 3111-3117 (2012).
33. Cahangirov S, Sahin H, Le Lay G, et al. Springer International Publishing: 63-85 (2017).
34. Song, Yu-Ling, et al. Applied Surface Science 256.21: 6313-6317 (2010).
35. Guo, Hongyan, et al. The Journal of Physical Chemistry C 118.25: 14051-14059 (2014).
36. Z Li, H Qian, J Wu, Gu BL, W Duan. Phys. Rev. Lett. 100(20):206802 (2008).
37. Kang, J., et al. Applied Physics Letters 100(23): 233122 (2012).
38. Cao, C., et al. Journal of Nanomaterials 2015: 1-10 (2015)..
39. Yang, Z., et al. Organic Electronics 14(11): 2916-2924 (2013).